# Shock Wave Structure for Argon, Helium, and Nitrogen

T.G. Elizarova[1], I.A. Shirokov[2], and S. Montero[3]

[1] *Institute of Mathematical Modeling, Russian Academy of Sciences, Miusskaya sq. 4a, Moscow, 125047, Russia*
[2] *Faculty of Computing Mathematics and Cybernetics, M. V. Lomonosov Moscow State University, Vorobjovy Gory, Moscow, 119899, Russia*
[3] *Instituto de Estructura de la Materia, CSIC, Serrano 121, Madrid, 28006, Spain*

**Abstract.** We compare the thickness of shock wave fronts at different Mach numbers, modeled via Navier-Stokes (NS) and Quasi-gasdynamic (QGD) equations, with experimental results from the literature. Monoatomic argon and helium, and diatomic nitrogen, are considered. In this modeling a finite-difference scheme with second-order spatial accuracy is employed. For argon the density thickness calculated via QGD and NS models are in good agreement with each other, and with the experimental results. For helium QGD and NS results agree well with those from the bimodal model. For nitrogen, the QGD results are closer to the experimental data than NS results. The QGD-based algorithm converges to the steady state solution faster than the NS-based one.

## 1. INTRODUCTION

The strong gradients within a shock wave lead to a number of associated effects of scientific and technological relevance which have attracted the attention of theoreticians and experimentalists for a long time. Though the main features of shock waves are well understood, much remains to be done to command the prediction of their quantitative aspects. The structure of a stationary shock wave has often been employed as testing problem for numerical models of rarefied gas flows. In this context the one-dimensional (1D) shock waves produced in atomic argon and helium, and in molecular nitrogen have been usual test systems to check the numerical aspects of Navier-Stokes (NS) equations.

The aim of this paper is to gain some understanding about the behavior of the equations underlying the 1D shock wave problem, widening the scope of NS-equations to the more general frame of the recently developed Quasi-gasdynamic (QGD) equations. The numerical results of both methods are compared with experiment for the test systems considering atomic argon and helium, and molecular nitrogen, where possible.

From the work done around 1960–1965 it was claimed that the NS approach could yield a reliable description of the density profile of the shock waves just up to Mach number $Ma \sim 2$. Their limitation to account for the shock wave thickness in $Ma > 2$ shock waves has been reported in detail by Kogan [1] and others [2–4]. Moreover, due to the sparse experimental temperature and velocity data on well-defined 1D shock waves produced in shock tubes, little was known about the actual merits of NS-equations to model the profiles of these quantities across the shock wave. This limitation has prevailed up to now in spite of the wealth of temperature data on 2D shock waves produced in jets [5, 6]. Unfortunately, 2D flows pose a number of difficulties for numerical modeling of shock waves structure (complicated flow pattern, low temperature upstream of the shock, with ill known dependence of the viscosity on the temperature). Therefore, 2D shock waves are little suited to be used in connection with the simple 1D mathematical formulation.

The main source of experimental 1D shock wave density data suited for the present purpose has been compiled, and completed with accurate original data by Alsmeyer [7]. In this work the experimental reciprocal thickness of argon and nitrogen 1D shock waves is shown to be well characterized up to $Ma = 10$, while a few sparse helium data points have been included. Complete argon and nitrogen experimental density profiles were also reported by Alsmeyer for several $Ma < 10$ values. Due to the shortage of experimental data on helium, the results of bimodal calculations have been also used as a reference [8], since they are known to be suitable for shock wave modeling.

On the other hand QGD calculations have been carried out before for the 1D shock problem, e.g. [9–12]. Their results have been compared with those from other numerical approaches, namely NS and DSMC, and also with some kinetic models.

In the present work the calculations of 1D shock wave structures have been carried out employing the QGD and the NS equations. The results are compared with experimental density measurements for argon, helium and nitrogen [7], as well as with results from the bimodal model [8]. Neither temperature nor velocity profiles were measured in the former experiments. According to the referred experimental data [7] the temperature upstream of the shock wave is 300 K, which allows us to employ a simple and well founded description of the viscosity with the parameters reported by Bird [13, App. I]. Other molecular properties of gases are also taken from [13].

## 2. MATHEMATICAL MODEL

The QGD equations for the one-dimensional plane flow read [14]:

$$\frac{\partial r}{\partial t} + \frac{\partial j_m}{\partial x} = 0, \qquad (1)$$

$$\frac{\partial (ru)}{\partial t} + \frac{\partial (j_m u)}{\partial x} + \frac{\partial p}{\partial x} = \frac{\partial \Pi_{xx}}{\partial x}, \qquad (2)$$

$$\frac{\partial E}{\partial t} + \frac{\partial (j_m H)}{\partial x} + \frac{\partial q}{\partial x} = \frac{\partial (\Pi_{xx} u)}{\partial x}. \qquad (3)$$

Here $r$ stands for the gas density, $u$ for the velocity, $p = rRT$ for the pressure, $T$ for the temperature, $g$ for the specific heat ratio, and $R$ for the gas constant; $E = ru^2/2 + p/(g-1)$ and $H = (E+p)/r$ are, respectively, the total energy and enthalpy per unit volume. The mass flux density vector is $j_m = r(u-w)$, where

$$w = \frac{t}{r}\frac{\partial}{\partial x}(ru^2 + p).$$

The $xx$ component of the shear-stress tensor employed in Eqs. (2) and (3), is given by

$$\Pi_{xx} = \frac{4}{3}h\frac{\partial u}{\partial x} + ut\left(ru\frac{\partial u}{\partial x} + \frac{\partial p}{\partial x}\right) + t\left(u\frac{\partial p}{\partial x} + gp\frac{\partial u}{\partial x}\right).$$

The heat flux vector is

$$q = q^{NS} - tru\left[\frac{u}{g-1}\frac{\partial}{\partial x}\left(\frac{p}{r}\right) + pu\frac{\partial}{\partial x}\left(\frac{1}{r}\right)\right], \text{ where } q^{NS} = -k\frac{\partial T}{\partial x}.$$

Viscosity coefficient $h$, heat conductivity coefficient $k$, and relaxation parameter $t$ are connected by the relations: $h = h_\infty (T/T_\infty)^W$, $k = gRh/((g-1)\cdot \text{Pr})$, $t = h/(p\cdot Sc)$, where $h_\infty$ is the value of $h$ for the temperature $T_\infty$; Pr and $Sc$ are Prandtl and Schmidt's numbers, respectively. For $t = 0$ the QGD system reduces to the NS system.

For the numerical solution of the system (1)–(3) the computational domain is covered with an uniform computational grid with the space step equal to $h_x$, and the time step equal to $h_t$. The space derivatives are approximated by the central differences of the second order. The time derivatives are approximated by forward differences of the first order. The finite-difference scheme for the initial-boundary problem (1)–(3) is solved by means of an explicit algorithm where the steady-state solution is attained as the limit of a time-evolving process. The space step is taken much smaller than the shock thickness. This ensures the stability of the numerical algorithm without resorting to any artificial dissipation.

We calculate the shock-wave structure for Mach numbers in the range 1.5–10. Dimensionless quantities are introduced on the basis of the upstream gas parameters. The initial conditions are the following: for $x < 0$, $r = r^{(1)} = 1$, $u = u^{(1)} = Ma$, $p = p^{(1)} = 1/g$, and for $x > 0$, the values of $r^{(2)}$, $u^{(2)}$, $p^{(2)}$ are defined via the Rankine–Hugoniot conditions. The values on the right and left boundaries are fixed. The numerical solution is supposed to be achieved according to the criterium $\max(\hat{r} - r)/h_t < e = 10^{-3}$.

The reciprocal shock thickness is defined as $l/d$, where $l$ is the upstream mean free path, and $d$ is the shock thickness, calculated in dimensionless form by the maximum value of the finite-difference derivative $\partial r/\partial x$:

$l/d = \max_i((r_{i+1} - r_{i-1})/2h_x)/(r^{(2)} - r^{(1)})$.

According to [13] the upstream mean free path is calculated as: $l = h/(r\sqrt{2pRT} \cdot \Omega/4)$, where $\Omega = 30/((7 - 2w)(5 - 2w))$.

## 3. RESULTS FOR ARGON AND HELIUM

The molecular parameters, taken from [13] are, for argon: $g = 5/3$, $w = 0.81$, $Sc = 0.752$, $\Pr = 2/3$. For helium: $g = 5/3$, $w = 0.66$, $Sc = 0.7575$, $\Pr = 2/3$. We use the QGD system (1)–(3), and also the NS system that can be obtained from (1)–(3) assuming $t = 0$. The number of space grid points is $N_x = 1200$, while the space step is $h_x = 0.25$. The time step is defined as $h_t = ah_x/\max(\sqrt{T} + |u|)$, $a = 0.001$.

The calculated reciprocal shock thickness for argon are shown on Fig.1 together with the experimental data collected in [7]. The results obtained by the author of [7] are marked by $\nabla$. One can see that the QGD and NS models yield close results for the shock wave thickness in a wide range of Mach numbers. These numerical results agree well with the experiment. However, the NS model demands about 5-10 times more time steps for convergence.

Fig. 2 shows the density, temperature, and velocity profiles in a $Ma = 9$ shock wave in argon. The density profiles shown are normalized as $f_r = (r - r^{(1)})/(r^{(2)} - r^{(1)})$, where $f_r$ is the shown value, and $r^{(1)}$, $r^{(2)}$ are the boundary values. A similar normalization holds for temperature, and $f_u = (u - u^{(2)})/(u^{(1)} - u^{(2)})$ for velocity. The experimental values for the density profile are taken from [7]. Both QGD and NS systems describe properly the density profile. For the positive values of $x$ the experimental data lie between NS and QGD profiles.

Fig. 3 represents the detailed density profile in argon for $Ma = 9$. Markers correspond to the computational points. While the QGD algorithm shows a smooth solution, one can notice the fluctuations of the NS-based numerical solution behind a shock wave. The period of the fluctuations equals the spatial grid step. These fluctuations explain the slow convergence of the NS numerical method in comparison with the QGD algorithm.

The employed numerical algorithm was tested for grid convergence. For this purpose the QGD-based modeling of the argon shock wave was carried out using a twice as fine space grid for $Ma = 10$, $a = 0.0001$, $N_x = 2400$, space step $h_x = 0.125$. In this variant we obtained $l/d = 0.2211561$, close to the reference grid result (0.220253). This assures that the results do not depend on the grid step size.

The results for helium are presented on Fig. 4, jointly with the few available experimental data on helium (from [7]) and the results from the bimodal approach [8]. The bimodal approach is reputed to describe well a shock wave structure up to high Mach numbers.

As in the case of argon, QGD and NS models yield for helium very close results for the shock thickness, with the QGD results somewhat closer to the reference data. The present results match well the experimental data for $Ma = 3$ and $Ma = 4$, and agree reasonably well with the bimodal calculation at higher $Ma$. As in previous calculations, the NS algorithm demands 5–10 times more time steps to convergence than the QGD approach.

## 4. RESULTS FOR NITROGEN

The following molecular parameters [13] were employed for nitrogen: $g = 7/5$, $w = 0.74$, $\Pr = 14/19$, $Sc = 0.746$. The computations were carried out using the QGD equations (1)–(3) and the NS equations. In order to accounting for the nonequilibrium between translational $T_t$ and rotational $T_r$ temperatures the QGDR system for diatomic gases was used [11] instead of the plain QGD system. The QGDR equations were used before for the calculation of shock structures in nitrogen, showing a good accordance with the kinetic calculations for 1D shock structure[11], and with the experimental results for underexpanded axisymmetric jets [6].

In present paper (unlike in [11]), the rotational relaxation time $t_r$ is calculated as $t_r = Z t_c$, $t_c = t(7-2w)(5-2w)/30$, where $Z = Z^\infty / [1 + (p^{3/2}/2)(T^*/T_t)^{1/2} + (p + p^2/4)(T^*/T_t)]$, $Z^\infty = 23$, $T^* = 91.5\ K$, $T^{(1)} = 273\ K$ [12]. The numerical algorithm for the QGDR equations is similar to the one described above and has been presented in [11].

The results for nitrogen are represented in Fig. 5 jointly with the experimental data from [7]; the data obtained by the author [7] are marked by $\nabla$. For Mach numbers between 1.5 and 3 the QGD and the NS results are similar. For higher Mach numbers the reciprocal shock wave thickness computed via NS model exceeds the QGD results. As well as in the previous calculations, the NS solution converges much slower then the QGD one. For $Ma \geq 7$ the numerical NS solution did not converge anymore. However, the QGD algorithm achieves the steady-state solution for all considered Mach numbers. In turn, the QGDR results show a much better agreement with experiment.

Fig. 6 shows the QGD, QGDR, and NS calculated density profiles, jointly with the experimental data for $Ma = 6.1$. For $x > 0$ the QGD density profile agrees better with the experimental data (taken from [7]) than the QGDR and NS profiles. For $x < 0$ the QGDR model yields the best solution.

## 8. CONCLUSIONS

In this work it is concluded that the QGD calculated density profiles and reciprocal shock thickness in argon agree well with experiment [7] for Mach numbers from 1.5 up to 10. The reciprocal shock thickness in helium agree well with the results of bimodal calculations [8], and with the few available experimental data. For diatomic nitrogen the QGD calculation overestimates the reciprocal shock thickness, compared with experiment. On the other hand, while the density profile is reasonably well described by the QGD equations, the reciprocal shock thickness is much better described by the QGDR system. This latter model takes into account the translational-rotational local nonequilibrium.

The QGD and NS calculations where performed under identical algorithmic conditions, employing the second order space approximation on a uniform spatial grid without artificial dissipation to stabilize the numerical solution. Both calculations yield similar final results, (but QGD are slightly more accurate than NS results). However the numerical results from NS equations converge about one order of magnitude slower than the QGD results. This is related with the numerical oscillations of NS solution. Such effect impairs attaining a converged NS solution, especially for large Mach numbers. For nitrogen a converged NS solution was obtained only up to $Ma = 6.1$. The number of time steps for convergence varies in a range of $10^5 - 10^8$.

In opposition to a widespread opinion, the NS equations prove to be precise enough for describing the density profile in an argon shock wave at Mach numbers $1.5 < Ma < 10$. The unreal limitation of NS equations for the shock wave problem at large Mach numbers may be attributed to numerical problems and imperfections of earlier calculations. See for instance Ref. [1–4], where the NS results for argon density thickness are off by ~ 100% at $Ma \sim 6$.

## REFERENCES

1. Kogan M. N, Rarefied Gas Dynamics, Plenum Press, New York, 1969.
2. Robben F., and Talbot L., Measurements of Shock Wave Thickness by the Electron Beam Fluorescence Method, Phys. Fluids, V. 9, N 4, pp. 633-643 (1966).
3. Toba K., and Melnik J. D., Two-Fluid Model For Shock Wave Structure, Phys. Fluids, V. 8, N 12, pp. 2153-2157 (1965).
4. Linzer M., and Hornig D. F., Structure of Shock Fronts in Argon and Nitrogen, Phys. Fluids, V. 6, N 12, pp. 1661-1668 (1963).

5. Ramos A., Maté B., Tejeda G., Fernández J. M., and Montero S., Raman Spectroscopy of Shock Waves, Phys. Rev. E, V. 62, N 4, pp. 4940-4945 (2000).
6. Graur I. A., Elizarova T. G., Ramos A., Tejeda G., Fernández J. M., and Montero S., Study of Shock Waves in Expanding Flows on the Basis of Spectroscopic Experiments and Quasi-Gasdynamic Equations, J. Fluid Mech., V. 504, pp. 239-270 (2004).
7. Alsmeyer H., Density Profiles in Argon and Nitrogen Shock Waves Measured by the Absorption of an Electron Beam, J. Fluid. Mech, V. 74, pp. 479-513 (1976).
8. Muckenfuss C., Some Aspects of Shock Structure According to Bimodal Model, Phys. Fluids, V. 5, N 11, pp. 1325-1336 (1962).
9. Elizarova T. G., Chetverushkin B. N., and Sheretov Yu. V., Quasi-Gasdynamic Equations and Computer Simulation of Viscous Gas Flows, Lecture Notes in Phys., N 414, Proc. 13$^{th}$ Intern. Conf. on Numer. Meth. in Fluid Dynamics, Roma, Springer-Verlag, 1992, pp. 421-425.
10. Elizarova T. G., Graur I. A., and Sheretov Yu. V., Quasi-Gasdynamic Equations and Computer Simulation of Rarefied Gas Flows, Proc. 19$^{th}$ Int. Symp. on Shock Waves, Marseille, France, July 26-30, 1993, Springer, V. 4, pp. 45-50.
11. Chirokov I. A., Elizarova T. G., and Lengrand J. C., Numerical Study of Shock Wave Structure Based on Quasi-Gasdynamic Equations With Rotational Nonequilibrium, 21$^{st}$ Int. Symp. on Rarefied Gas Dynamics, Ed. R.Brun et al., Cepadues, Toulouse, France, 1999, V. 1, pp. 175-182.
12. Elizarova T. G, Graur I. A., and Lengrand J.-C., Macroscopic Equations for a Binary Gas Mixture, 22$^{nd}$ Int. Symp. on Rarefied Gas Dynamics, Ed. T. G. Bartel, M. A. Gallis, AIP Conference Proceedings, V. 585, Melville, New York, 2001, pp. 297-304.
13. Bird G. A., Molecular Gas Dynamics and the Direct Simulation of Gas Flows, Clarendon press, Oxford, 1998.
14. Elizarova T. G., and Sheretov Yu. V., Theoretical and Numerical Analysis of Quasi-Gasdynamic and Quasi-Fluid-dynamic Equations, Comput. Math. and Math. Phys., V. 41, N 2, pp. 219-234 (2001).

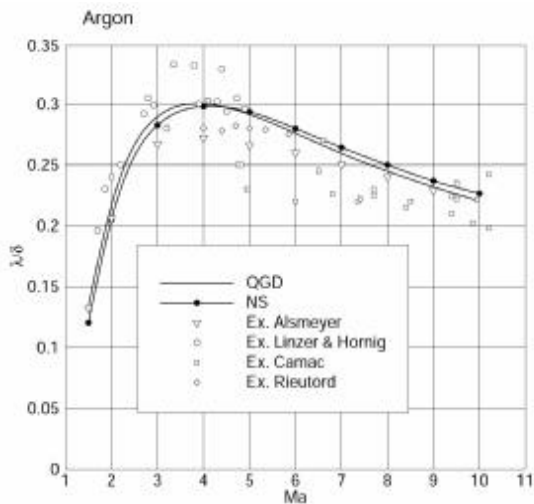
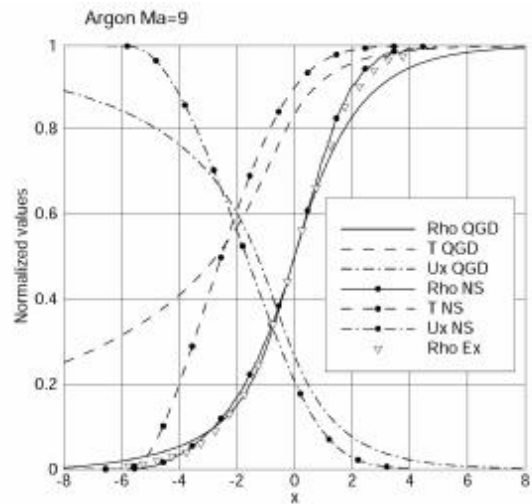

**FIGURE 1.** Reciprocal shock wave thickness versus Mach number in argon

**FIGURE 2.** Density, velocity and temperature profiles in argon shock wave for $Ma = 9$

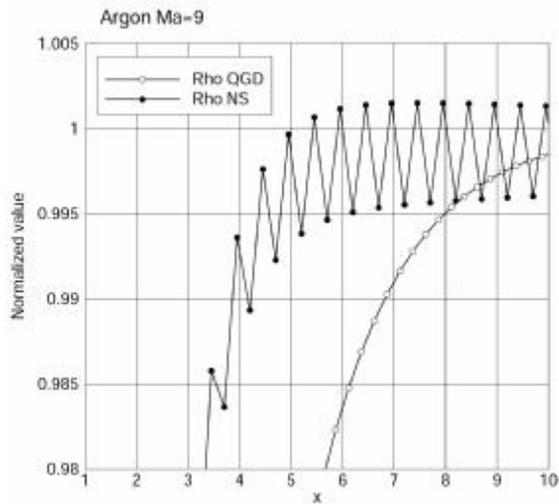

**FIGURE 3.** Density profiles in argon shock wave for $Ma = 9$ in detail

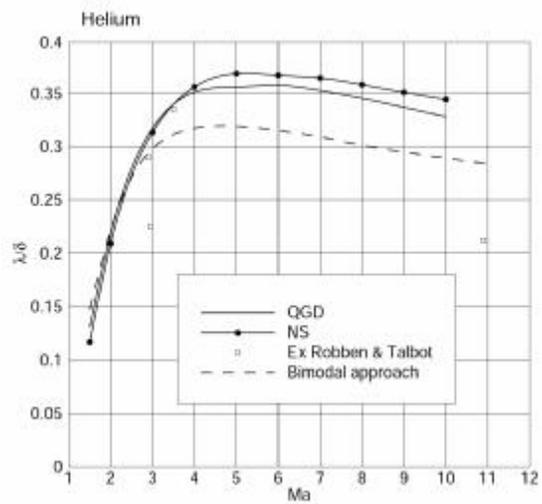

**FIGURE 4.** Reciprocal shock wave thickness versus Mach number in helium

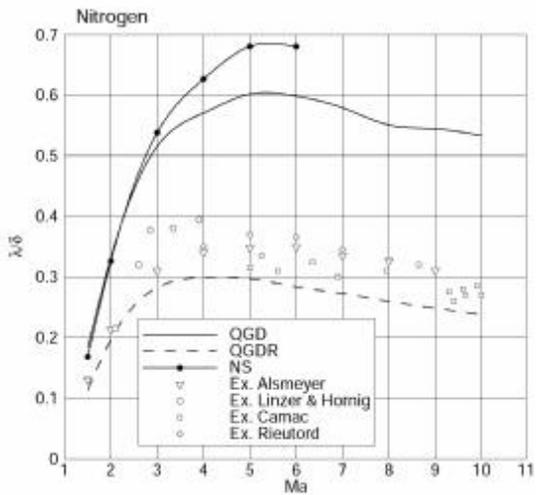

**FIGURE 5.** Reciprocal shock wave thickness versus Mach number in nitrogen

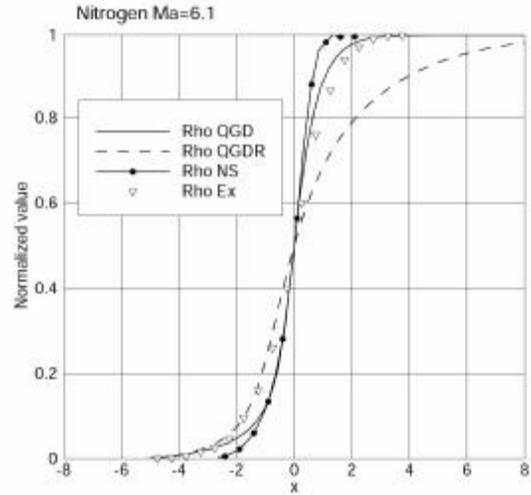

**FIGURE 6.** Density profiles in nitrogen shock wave for $Ma = 6.1$